\documentclass[pra,twocolumn,superscriptaddress]{revtex4-1}
\usepackage{amssymb,amsmath,graphicx,color}
\usepackage{times}
\usepackage{bm}
\usepackage{appendix}
\usepackage{xcolor}
\usepackage[colorlinks,citecolor=blue]{hyperref}
\usepackage{mathrsfs}
\usepackage{dcolumn}% Align table columns on decimal point
\usepackage{float}
\usepackage{array}

\begin{document}

\def\sgn{\mathop{\rm sgn}}

\title{Detecting non-Abelian statistics of topological states on a chain of superconducting circuits}

\author{Jun-Yi Cao}

\author{Jia Liu}  \email{liuj.phys@foxmail.com}
\affiliation{Guangdong Provincial Key Laboratory of Quantum Engineering and Quantum Materials,
GPETR Center for Quantum Precision Measurement, and School of Physics and Telecommunication Engineering,
	South China Normal University, Guangzhou 510006, China}

\author{L. B. Shao}
\email{lbshao@nju.edu.cn}
\affiliation{National Laboratory of Solid State Microstructures and Department of Physics, Nanjing University, Nanjing 210093, China}

\author{Zheng-Yuan Xue}
\email{zyxue83@163.com}
\affiliation{Guangdong Provincial Key Laboratory of Quantum Engineering and Quantum Materials,
GPETR Center for Quantum Precision Measurement,
and School of Physics and Telecommunication Engineering,
	South China Normal University, Guangzhou 510006, China}

\date{\today}

\begin{abstract}
 In view of the fundamental importance and many promising potential applications, non-Abelian statistics of topologically protected states have attracted much attention recently. However, due to the operational difficulties in solid-state materials, experimental realization of non-Abelian statistics is lacking. The superconducting quantum circuit system is scalable and controllable, and thus is a promising platform for quantum simulation. Here we propose a scheme to demonstrate non-Abelian statistics of topologically protected zero-energy edge modes on a chain of superconducting circuits. Specifically, we can realize topological phase transition by varying the hopping strength and magnetic field in the chain, and the realized non-Abelian operation can be used in topological quantum computation. Considering the advantages of the superconducting quantum circuits, our protocol may shed light on quantum computation via topologically protected states.
\end{abstract}

\maketitle

\section{Introduction}
Following Feynman's  suggestion about the possibility of a quantum computer, Shor proposed a quantum algorithm that could efficiently solve the prime-factorization problem \cite{QC,petershor}. Since then the research of quantum computation has become controversial. Recently, topological quantum computation has become one of the perfect constructions to build a quantum computer. The protocols based on the topological systems are built by neither bosons nor fermions, but so-called non-Abelian anyons, which obey non-Abelian statistics. Therefore, the realization that particles obey non-Abelian statistics in different physical systems has led to wide-ranging research. Physical systems with the fractional quantum Hall effect have been developed extensively as a candidate for topological quantum computation and similarly Majorana fermions also have attracted a great deal of attention in related research \cite{natphys,Kitaev_anyon,FQH_FuKane,FQH_Nagosa,FQH_DasSarma,FQH_Alicea, FQH_Fujimoto,FQH_SCZhang,wenxg_prl,kitaev_prl,read_prb,adystern_nature,DasSarma_RMP, Kitaev_majorana,DarSarma_majorana,liu_fop,science_majorana}. However, experimental non-Abelian operations are still being explored for real quantum computation, and thus the relevant research still has great significance.

Recently, the superconducting quantum circuit system \cite{cqed1,cqed2,cqed3,Nori-rew-Simu2-JC}, a scalable and controllable platform which is suitable for quantum computation and simulation \cite{s1,s2,s3,s4,s5,s6,s7,s8,s9,TS1D_prl}, has attracted a great deal of attention and has been applied in many studies. For example, the Jaynes-Cummings (JC) model \cite{JCModel}, describing the interaction of a single two-level atom with a quantized single-mode photon, can be implemented by a superconducting transmission line resonator (TLR) coupled with a transmon. Meanwhile, JC units can be coupled in series, by superconducting quantum interference devices (SQUIDs), forming a chain \cite{Gu} or a two-dimensional lattice \cite{xue,liu_QSH}, providing a promising platform for quantum simulation and computation. Compared with cold atoms and optical lattice simulations \cite{SOC1,SOC2,SOC3}, the superconducting circuits possess good individual controllability and easy scalability.

Here we propose a scheme to demonstrate non-Abelian statistics of topologically protected zero-energy edge modes on a chain of superconducting circuits. Each site of the chain consists of a JC coupled system, the single-excitation manifold of which mimics spin-1/2 states. Different neighboring sites are connected by SQUIDs. In this setup, all the on-site potential, tunable spin-state transitions, and synthetic spin-orbit coupling can be induced and adjusted independently by the drive detuning, amplitude, and phases of the AC magnetic field threading through the connecting SQUIDs. These superconducting circuits also have been used in previous work \cite{Gu}, which we will compare with our work in the last paragraph of Sec.III. With appropriate parameters, topological states and the corresponding non-Abelian statistics can be explored and detected.

\section{The model}
\subsection{The proposed model}
We propose to implement non-Abelian quantum operations in a one-dimensional (1D) lattice with the Hamiltonian
\begin{equation}
\begin{aligned}
H&=\sum_l t_z(c_{l,\uparrow}^\dag c_{l+1,\uparrow}-c_{l,\downarrow}^\dag c_{l+1,\downarrow})+\text{H.c.}\\
&+\sum_l h_z(c_{l,\uparrow}^\dag c_{l,\uparrow}-c_{l,\downarrow}^\dag c_{l,\downarrow})\\
&-\sum_l i\Delta_0 e^{-i\varphi}(c_{l,\uparrow}^\dag c_{l+1,\downarrow}-c_{l+1,\uparrow}^\dag c_{l,\downarrow})+\text{H.c.},
\label{eq4}
\end{aligned}
\end{equation}
where $c^{\dag}_{l,\alpha}=|\bar{\alpha}\rangle_l \langle G|$ and $c_{l,\alpha}=|G\rangle_l \langle \bar{\alpha}|$ are the creation and annihilation  operators for a polariton with ``spin'' $\alpha$ in $l$th unit cell, respectively.
First, If we set $\varphi = 0$, the Hamiltonian in Eq. (\ref{eq4}) can be transformed to the momentum space as $H=\sum_{{\bf{k}}}\Psi^{\dag}_{\bf{k}}\hat{h}({\bf{k}})\Psi_{\bf{k}}$, where $\Psi_{\bf{k}}=\left( c_{{\bf{k}},\uparrow}, c_{{\bf{k}},\downarrow} \right)^\intercal$ and
\begin{equation}
\hat{h}({\bf{k}})=\left[h_z+2t_z\cos({\bf{k}})\right]\sigma_z+ 2\Delta_0\sin({\bf{k}})\sigma_x,
\label{eq5}
\end{equation}
with lattice spacing $a = 1$ and Pauli matrices $\sigma_x$ and $\sigma_z$. The energy bands of this system are given as
\begin{equation}
E({\bf{k}}) = \pm\sqrt{ \left[h_z+2t_z\cos({\bf{k}})\right]^2+ \left[2\Delta_0\sin({\bf{k}})\right]^2},
\label{energy_band}
\end{equation}
which indicates that the energy gap will close only when $h_z = \pm2t_z$. It is well known that a topological phase transition occurs when the gap closes and opens. In order to identify the topological zero-mode states $\psi_{0}$, we start from a semi-infinite chain. There is a chiral symmetry $\sigma_y\hat{h}({\bf{k}})\sigma_y = -\hat{h}({\bf{k}})$, so this system belongs to the topological class AIII with topological index $\mathbb{Z}$ \cite{topoclass}. The topological invariant of this system is relevant to the so-called non-Abelian charge, which represents that the system obeys the non-Abelian statistics \cite{topo-nonablian,nonabliancharge}. If there is a $\psi_{0}$ state inside the gap, $\sigma_y\psi_{0}$ is identical to $\psi_{0}$ up to a phase factor, since that under the chiral symmetry $E({\bf{k}}) \rightarrow -E({\bf{k}})$. As a result, $\psi_{0}$ must be an eigenstate of $\sigma_y$ as $\phi_{\pm}=\frac{1}{\sqrt{2}} (1, \pm i)^\intercal$ and
\begin{equation}
h_z+2t_z\cos({\bf{k}})=\mp 2i\Delta_0\sin({\bf{k}}).
\label{eq6}
\end{equation}
These two equations are obtained by substituting $\phi_{\pm}$ into Eq. (\ref{eq5}), which is necessary to satisfy the former conditions. Note that $\sigma_x\phi_{\pm} = \pm i\phi_{\mp}$, $\sigma_y\phi_{\pm} = \pm \phi_{\pm}$, and $\sigma_z\phi_{\pm} = \phi_{\mp}$; according to Eq. (\ref{eq5}), $\Delta_0 \rightleftharpoons -\Delta_0$ is equivalent to $\hat{h}({\bf{k}}) \rightarrow \sigma_z \hat{h}({\bf{k}})\sigma_z$, $h_z = 0$, and $t_z \rightleftharpoons -t_z$ is equivalent to $\hat{h}({\bf{k}}) \rightarrow \sigma_x \hat{h}({\bf{k}}) \sigma_x$.

\begin{figure}[tb]		
\centering
\includegraphics[width=0.88\columnwidth]{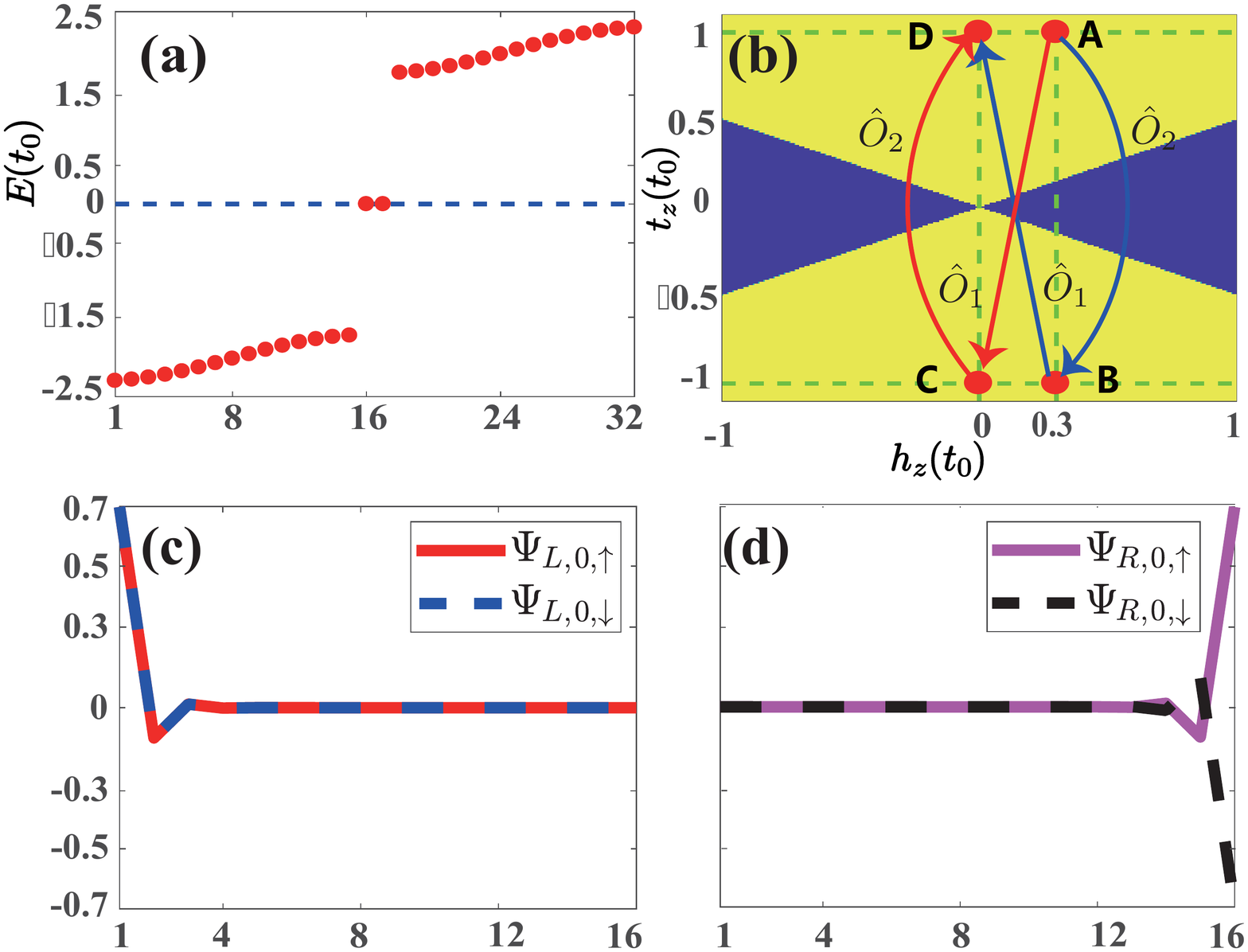}
\caption{Numerical calculations in 1D, consisting 16 unit-cell lattices. (a) Eigenenergies of the finite system with the parameters $t_z/t_0=1$, $\Delta_0/t_0 = 0.99$, and $h_z/t_0 = 0.3$. An energy gap is opened in the bulk and there are two zero modes in the middle of the gap. It is obvious that those two zero modes are localized at different edges. (b) The Phase diagram of the chain with arbitrary $\Delta_0$. The yellow and dark blue regions show the topologically invariant $\nu = 1$ and $\nu = 0$, respectively. Four red circles A, B, C, and D represent the parameters for demonstrating the non-Abelian quantum transformations. The red arrow means that $\hat{O}_1$ is executed first and then $\hat{O}_2$ follows; the blue arrow means that $\hat{O}_2$ is executed first and then $\hat{O}_1$. (a), (c), and (d) With the same parameters, the left and right zero-energy edge states of the chain are numerically calculated, where the red or purple line plots the amplitude of the $|$$\uparrow\rangle$ components of the wave functions and the blue or black dashed line plots the amplitude of the $|$$\downarrow\rangle$ components of the wave functions divided by a phase factor $i$.}
\label{figure1}
\end{figure}

Since there is a gap in the bulk, these equations only have complex solutions which provide localized states at the edges. In order to satisfy the boundary conditions  $\psi_{0}|_{x=0} = 0$ and $\psi_{0}|_{x=\infty} = 0$, the solutions of Eq. (\ref{eq6}) for the same eigenstate $\phi_{\pm}$ must satisfy $\mathbf{Im}({\bf{k}}) > 0$, as there is no superposition of orthogonal states to satisfy this vanishing boundary condition. Careful analysis shows that there is an edge state $\phi_+$ localized at $x = 0$ when $|h_z| < 2t_z$, $t_z > 0$, and $\Delta_0 > 0$.

\subsection{Non-Abelian statistics}
In order to set up a scheme that can be achieved in experiments, we consider a chain with a finite number of cells. Fortunately, topologically protected zero modes are stable until the energy gap is closed and thus can survive under local perturbations, a robust quantum computation can be realized using those modes. For a finite system, the same argument can be applied to the edge states. After numerical calculations, we find that a chain with 16 lattices is sufficient to realize a non-Abelian operation with corresponding parameters. In all the following numerical calculations, we set $\varphi=0$ and $t_0$ is the energy unit. We set the energy levels of the Hamiltonian in Eq. (\ref{eq4}) with the parameters $t_z/t_0 = 1$, $\Delta_0/t_0 = 0.99$, and $h_z/t_0 = 0.3$. As shown in Fig. \ref{figure1}(a), we can find zero-energy modes that can be used to demonstrate their non-Abelian statistics.

We choose 4 such modes to realize non-Abelian operation in our scheme; these are the four circles dots in Fig. \ref{figure1}(b). In addition, we also calculate the topological invariants \cite{Chern,numbers,invariants,Xiao}
\begin{equation}
\begin{aligned}
\nu = \frac{1}{2}\left[\sgn(2t_z+h_z)+\sgn(2t_z-h_z) \right], t_z>0\notag\\
\nu = -\frac{1}{2}\left[\sgn(2t_z+h_z)+\sgn(2t_z-h_z) \right], t_z<0.
\end{aligned}
\end{equation}
According to $\nu$, we divide the $t_z$-$h_z$ plane into topologically nontrivial and trivial phases, as in Fig. \ref{figure1}(b). We set two quantum operations $\hat{O}_1$ and $\hat{O}_2$, where $\hat{O}_1$ is implemented by first changing the signs of $t_z$ and $\Delta_0$ and then varying $h_z$ from $0.3t_0$ to 0 and $\hat{O}_2$ is obtained with constant $h_z$ while changing the signs of $t_z$ and $\Delta_0$. We choose the initial state as $|\Psi_{i}(x)\rangle = |\Psi_{L,0}(x)\rangle$ and calculate the edge states of the four parameters(red circles) related to non-Abelian quantum operations in Fig. \ref{figure1}(b), which are described in the following.

$Dot A$. When $t_z/t_0 = 1$, $\Delta_0/t_0 = 0.99$, and $h_z/t_0 = 0.3$, the two zero-mode edge states of the system are
\begin{eqnarray}\label{eq7}
|\Psi_{L,0}(x)\rangle&=N_0\frac{\left[ \left(\frac{-b_0+\sqrt{c_0}}{2}\right) ^x - \left(\frac{-b_0-\sqrt{c_0}}{2}\right) ^x \right]}{\sqrt{c_0}}\phi_+, \\
|\Psi_{R,0}(x)\rangle&=N_0\frac{\left[ \left(\frac{-b_0+\sqrt{c_0}}{2}\right) ^{N-x+1}- \left(\frac{-b_0-\sqrt{c_0}}{2}\right) ^{N-x+1} \right]}{\sqrt{c_0}}\phi_-,\notag
\end{eqnarray}
as shown in Figs. \ref{figure1}(c) and \ref{figure1}(d), where $a_0 = (t_0-\Delta_0)/(t_0+\Delta_0)$,  $b_0 = h_z/(t_0+\Delta_0)$,  $c_0 = b_0^2-4a_0$, $N$ is the number of cells, and $N_0$ is a normalized constant that can only be solved numerically.

$Dot B$. When $t_z \rightarrow -t_z$,  $\Delta_0 \rightarrow -\Delta_0$, and $h_z/t_0 = 0.3$, the two edge states can be obtained as
\begin{equation}
\begin{aligned}
|\Psi_{L,1}(x)\rangle&=N_1\frac{\left[ \left(\frac{b_1+\sqrt{c_1}}{2}\right) ^x - \left(\frac{b_1-\sqrt{c_1}}{2}\right) ^x \right]}{\sqrt{c_1}}\phi_+, \\
|\Psi_{R,1}(x)\rangle&=N_1\frac{\left[ \left(\frac{b_1+\sqrt{c_1}}{2}\right) ^{N-x+1}- \left(\frac{b_1-\sqrt{c_1}}{2}\right) ^{N-x+1} \right]}{\sqrt{c_1}}\phi_-.
\end{aligned}
\label{eq9}
\end{equation}
where $a_1 = 1/a_0$,  $b_1 = h_z/(t_0-\Delta_0)$,  $c_1 = b_1^2-4a_1$, and $N_1$ is a normalized constant that can only be solved numerically.

$Dot  C$. When $t_z \rightarrow -t_z$,  $\Delta_0 \rightarrow -\Delta_0$, and $h_z/t_0 = 0.3 \rightarrow h_z = 0$, the two edge states can be obtained as
\begin{equation}\label{eq8}
\begin{aligned}
|\Psi_{L,2}(x)\rangle&=N_2 \sin\left(\frac{\pi}{2} x \right) e^{-(a_2/2)x }  \phi_+, \\
|\Psi_{R,2}(x)\rangle&=N_2 \sin\left(\frac{\pi}{2} (N-x+1) \right) e^{-(a_2/2) (N-x+1) } \phi_-.
\end{aligned}\end{equation}
where $N_2 = \sqrt{2\sinh{a_2}}$ and $a_2 = \ln{(1/a_0)}$.

$Dot  D$. When $-t_z \rightarrow t_z$,  $-\Delta_0 \rightarrow \Delta_0$, and $h_z = 0$, the two edge states can be obtained as
\begin{equation}
\begin{aligned}
|\Psi_{L,3}(x)\rangle&=|\Psi_{L,2}(x)\rangle, \\
|\Psi_{R,3}(x)\rangle&=|\Psi_{R,2}(x)\rangle.
\end{aligned}
\label{eqP3}
\end{equation}

We now proceed to detail a demonstration of our non-Abelian statistics for the zero modes. Specifically, we show that changing the order of two operations $\hat{O}_1$ and $\hat{O}_2$ that are applied to an initial state $|\Psi_{i}(x)\rangle$ will lead to different final states. We consider the case in which the $\hat{O}_1$ operation is implemented first, which is equivalent $\phi_{+} \rightarrow \phi_{-}$ so that $\hat{O}_1|\Psi_{i}(x)\rangle = |\Psi_{R,2}(x)\rangle$. When $\hat{O}_2$ is applied to $|\Psi_{R,2}(x)\rangle$, the Hamiltonian experiences two unitary transformations $\sigma_x$ and $\sigma_z$, which are equivalent $\phi_{-} \rightarrow \phi_{+} \rightarrow \phi_{-}$, so we can get $\hat{O}_2|\Psi_{R,2}(x)\rangle =  |\Psi_{R,3}(x)\rangle = |\Psi_{f}(x)\rangle$. As a result, the initial state $|\Psi_{i}(x)\rangle$ passes through $\hat{O}_1$ and then passes through $\hat{O}_2$, eventually transforming into the final state $|\Psi_{f}(x)\rangle$, which corresponds to the directions of the two red arrows in Fig. \ref{figure1}(b), i.e., $|\Psi_{i}(x)\rangle \xrightarrow{\hat{O}_1} |\Psi_{R,2}(x)\rangle \xrightarrow{\hat{O}_2} |\Psi_{f}(x)\rangle$.

Alternatively, when  the quantum operation $\hat{O}_2$, i.e., $\phi_{+} \rightarrow \phi_{-}$, is applied to the initial state $|\Psi_{i}(x)\rangle$ first, we can get $\hat{O}_2|\Psi_{i}(x)\rangle = |\Psi_{R,1}(x)\rangle$. Then $\hat{O}_1$ is applied to $|\Psi_{R,1}(x)\rangle$, i.e., $\phi_{-} \rightarrow \phi_{+}$, and we can get $\hat{O}_1|\Psi_{R,1}(x)\rangle = |\Psi_{L,3}(x)\rangle = |\Psi^{'}_{f}(x)\rangle$. As a result, the initial state $|\Psi_{i}(x)\rangle$ passes through $\hat{O}_2$ and then passes through $\hat{O}_1$, eventually transforming into $|\Psi^{'}_{f}(x)\rangle$, which corresponds to the directions of the two blue arrows in Fig. \ref{figure1}(b), i.e., $|\Psi_{i}(x)\rangle \xrightarrow{\hat{O}_2} |\Psi_{R,1}(x)\rangle \xrightarrow{\hat{O}_1} |\Psi^{'}_{f}(x)\rangle$. The above operations could be written as 
\begin{equation}
\begin{aligned}
&\hat{O}_2\hat{O}_1|\Psi_{i}(x)\rangle=|\Psi_{f}(x)\rangle,
&\hat{O}_1\hat{O}_2|\Psi_{i}(x)\rangle=|\Psi^{'}_{f}(x)\rangle.
\label{operation}
\end{aligned}
\end{equation}

It can be seen that $|\Psi_{f}(x)\rangle$ and $ |\Psi^{'}_{f}(x)\rangle$ are different in the position distribution, and the two final states can be distinguished experimentally by measuring the position, which we will show later in Figs. \ref{figure3}(a) and \ref{figure3}(b).We note that the systematic parameters should be changed in an adiabatic way; the dynamical details of this process are discussed in Ref. \cite{natphys}. The adiabatic condition is justified in the following. In the initial state $|\Psi_i(x)\rangle$, we set the parameters $t_0 =2\pi \times 4$ MHz and $h_z/t_0 = 0.3$, which corresponds to an energy gap  $\Delta \simeq 2\pi \times 43$ MHz, and thus the diabatic evolution time scale is $\tau=1/\Delta\simeq23$ ns. In Figs. \ref{figure3}(a) and \ref{figure3}(b) we plot the two final-states changes over a time duration of $\emph{T} = 3 \mu$s and it can be seen that the edge state has almost no decay. Note that $T/\tau> 42$ and thus the adiabatic evolution condition of our system can be safely met.

\begin{figure}[tbp]		
\centering
\includegraphics[width=0.85\columnwidth]{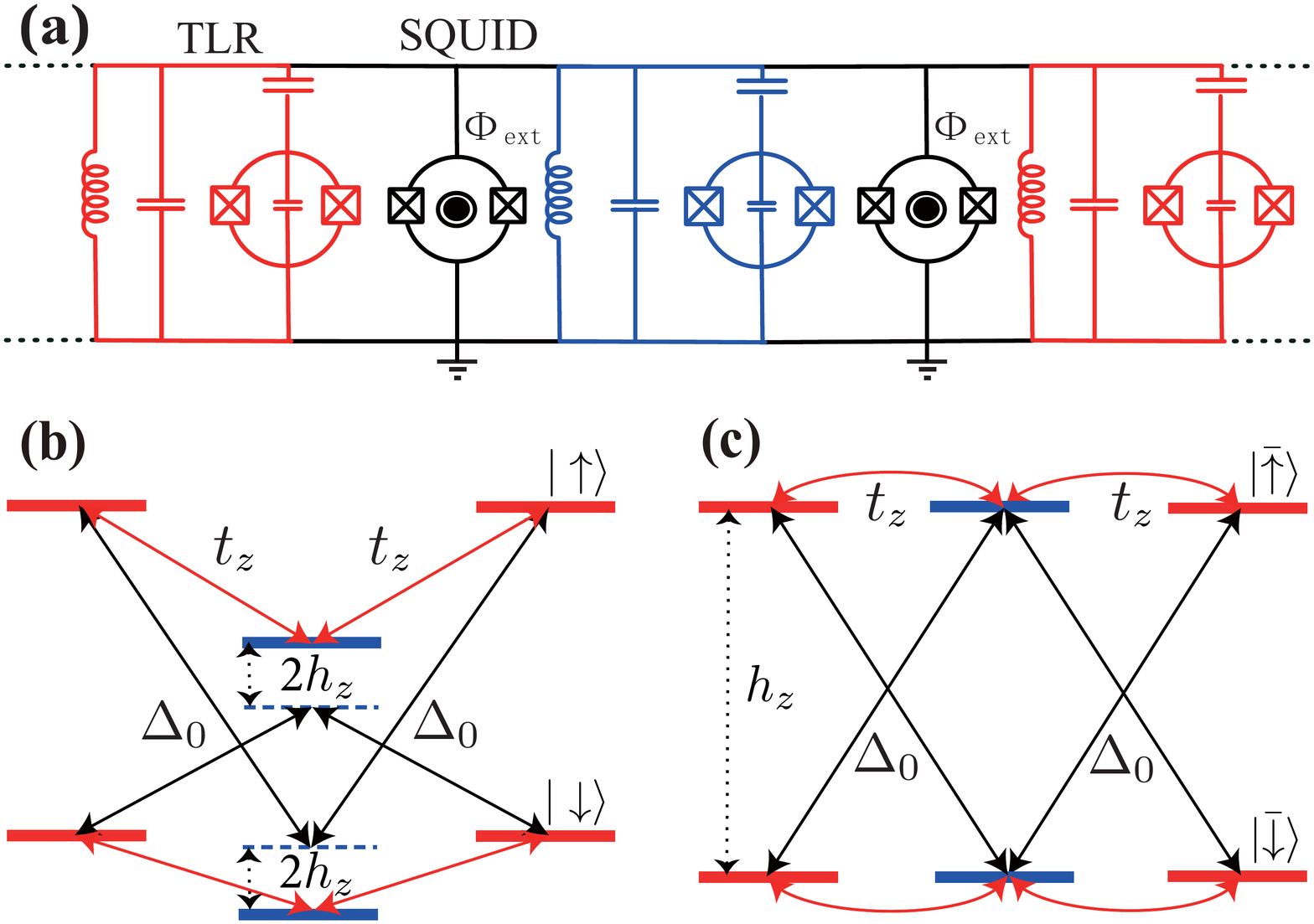}
\caption{Proposed setup of the superconducting circuit to mimic a spin-1/2 lattice model. (a) Spin-1/2 polariton lattice with two types of unit cells, R-type (colored red) and B-type (colored blue), arranged alternately, which are of a unit cell and are of different qubit and photon eigenfrequencies and JC coupling strengths. Each unit cell has two pseudo-spin-1/2 states simulated by the two single-excitation eigenstates of the JC model. The neighboring unit cells are coupled by a combination of a SQUID in series, to induce the tunable inter-cell photon hopping. (b) Detuned couplings of intercell spin states. In order to achieve the simulated Hamiltonian, two sets of driving strength are assigned: The coupling strength between $|\bar{\uparrow}\rangle_{l}\leftrightarrow |\bar{\uparrow}\rangle_{l+1}$ and $|\bar{\downarrow}\rangle_{l}\leftrightarrow |\bar{\downarrow}\rangle_{l+1}$ is $t_z$ (red) and the coupling strength between $|\bar{\uparrow}\rangle_{l}\leftrightarrow |\bar{\downarrow}\rangle_{l+1}$ and $|\bar{\downarrow}\rangle_{l}\leftrightarrow |\bar{\uparrow}\rangle_{l+1}$ is $\Delta_0$ (black). (c) Polariton lattice in a rotating frame, where all polariton lattices can be considered the same, so that the proposed circuit simulates a 1D spin-1/2 tight-binding lattice model.}
\label{figure2}
\end{figure}	

\section{Implementation}
%\subsection{Janeys-Cummings lattice}
 Following the previous discussion, now we will show how to realize our proposal in a superconducting circuit system. The method of realizing the 1D JC lattice in the superconducting circuit is shown in Fig. \ref{figure2}(a). We set the red and blue lattices alternately connected in a series on one chain. Each lattice contains a JC coupling, where a TLR and a transmon are employed with resonant interaction \cite{cqed1,Nori-rew-Simu2-JC}, and the adjacent lattices are connected by a grounded SQUID. As a result, setting $\hbar=1$ hereafter, the Hamiltonian of this JC lattice  is
\begin{equation}
H_{\text{JC}}=\sum_{l=1}^N h_l+\sum_{l=1}^{N-1} J_l(t)(a_l^\dag a_{l+1}+\text{H.c.}),
\label{eq1}
\end{equation}
where $N$ is the number of the unit cells and $h_l=\omega_l \sigma_l^\dag \sigma_l^{-}+\omega_l a_l^\dag a_l+g_l(\sigma_l^\dag a_l+\text{H.c.})$ is the JC-type interacting Hamiltonian in the $l$th unit cell, with $\sigma_l^\dag=$$|$$\text{e}\rangle_l\langle \text{g}$$|$ and $\sigma_l^{-}=$$|$$\text{g} \rangle_l\langle \text{e}$$|$ the raising and lowering operators of the $l$th transmon qubits, respectively, and $a_l$ and $a^\dag$ the annihilation and creation operators of the photon in the $l$th TLR, respectively. The condition $g_l \ll  \omega_l$ has to be met to justify the JC coupling. Its three lowest-energy dressed states are $|$$\uparrow\rangle_l=\frac{1}{\sqrt{2}}($$|$$\text{0e}\rangle_l+$$|$$\text{1g}\rangle_l), $$|$$\downarrow\rangle_l =\frac{1}{\sqrt{2}}($$|$$\text{0e}\rangle_l-$$|$$\text{1g}\rangle_l)$, and $|$$\text{0g}\rangle_l$, with the corresponding energies $E_{l,\uparrow}=\omega_l+g_l, E_{l\downarrow}=\omega_l-g_l$, and $0$. In addition, $J_l(t)$ is the intercell hopping strength between the $l$th and $(l+1)$th unit cells. Here we exploit the two single-excitation eigenstates $|$$\uparrow \rangle_l$ and $|$$\downarrow \rangle_l$ to simulate the effective electronic spin-up and spin-down states; they are regarded as a whole and are referred to as a polariton.

%\subsection{Polaritonic spin-orbit coupling}
We will show how the coupling strength $J_l(t)$ is regulated by regulating the magnetic flux of the adjacent TLRs and SQUIDs. Because two single-excitation dressed states act as two pseudospin states in each cell, there are four hoppings between two adjacent cells. To control the coupling strength and phase of each hopping, we introduce four driving field frequencies in each $J_l(t)$. For this purpose, we adopt two sets of unit cells, R type and B type, which are alternately linked on one chain [see Fig. \ref{figure2}(a)]. Setting the chain started with an R type unit cell, when $l$ is odd (even), $\omega_l=\omega_{\text{R}}(\omega_{_{\text{B}}})$ and $g_l=g_{\text{R}}(g_{\text{B}})$. Then we set $\omega_{\text{R}}/2\pi = 6$ GHz, $\omega_{\text{B}}/2\pi = 5.84$ GHz, $g_{\text{R}}/2\pi = 200$ MHz, and $g_{\text{B}}/2\pi =120 $ MHz. In this way, the energy interval of the four hoppings is $\{|E_{l,\alpha}-E_{l+1,\alpha'}|/2\pi\}_{\alpha, \alpha'=\uparrow /\downarrow}=\{ 80, 160, 240, 480\}$ MHz. The frequency distance between each pair is no less than 20 times the effective hopping strength $t_0/2\pi = 4$ MHz, with $t_z/t_0 = 1$ and $\Delta_0/t_0 = 0.99$, and thus they can be selectively addressed in terms of frequency. Then the driving $J_l(t)$ has to correspondingly contain four tunes, written as
 \begin{equation}
J_l(t)=\sum_{\alpha,\alpha'}4t_{l,\alpha,\alpha'} \cos\left(\omega_{l,\alpha,\alpha'}^dt+\varphi_{l,\alpha,\alpha'} \right),
\end{equation}
where $l=1,2,\ldots,N$ and $\alpha,\alpha'\in \{\uparrow,\downarrow\}$. We will show that the time-dependent coupling strength $J_l(t)$ can induce a designable spin transition under a certain rotation-wave approximation. First, we calculate the form of the Hamiltonian (\ref{eq1}) in the single-excitation state of the direct product space$\{|0\text{g}, \ldots , 0\text{g}, \underset{l\text{th}}{\alpha}, 0\text{g}, \ldots, 0\text{g} \rangle \}$. Hereafter, we use $|\bar{\alpha} \rangle_l$ to denote $|0\text{g}, \ldots, 0\text{g}, \underset{l\text{th}}{\alpha}, 0\text{g}, \ldots, 0\text{g}\rangle$, and $|G\rangle$ to denote $|0\text{g}, \ldots, 0\text{g}, \rangle$. Then we define a rotating frame by $U(t)=exp\left\{-i\sum_l\left[h_l-\sum_{\alpha}p_{l,\alpha}|\bar{\alpha} \rangle_l \langle \bar{\alpha}|\right]t\right\}$, where $\alpha=\uparrow,\downarrow$, $p_{l,\alpha}$ are parameters, which will be determined according to the Hamiltonian to be simulated. And map the Hamiltonian in Eq. (\ref{eq1}) into the single-excitation subspace span $\{|\bar{\alpha}\rangle_l\}$ to get
\begin{eqnarray} \label{eq16}
H'_{\text{JC}}&=&U^{\dag}H_{\text{JC}}U+i\dot{U}^{\dag}U\notag\\
&=&\sum_{l=1}^{N} \left[ \sum_{\alpha}p_{l,\alpha}|\bar{\alpha}\rangle_l \langle \bar{\alpha}|\right]+U^{\dag} \left(\sum_{l=1}^{N-1}h_{int}^l \right)U.
\end{eqnarray}
%where $\delta_{\alpha,\alpha'}$ is Dirac delta function.
Selecting $\omega_{l,\alpha,\alpha'}^d=(E_{l,\alpha}-p_{l,\alpha})-(E_{l+1,\alpha'}-p_{l+1,\alpha'})$, under the rotating-wave approximation, i.e., $|\omega_{l,\alpha,\alpha'}^d|\gg t_{l,\alpha,\alpha'}$ and $|\omega_{l,\alpha,\alpha'}^d\pm\omega_{l,\alpha'',\alpha'''}^d|\gg t_{l,\alpha'',\alpha'''}$,  Eq. (\ref{eq16})  is simplified to
\begin{eqnarray}\label{eq17}
&&H'_{\text{JC}}=\sum_{l=1}^{N} \sum_{\alpha}p_{l,\alpha}|\bar{\alpha}\rangle_l \langle\bar{\alpha}|\\
&&+\sum_l^{N-1} \sum_{\alpha,\alpha'}\{ t_{l,\alpha,\alpha'} (2\delta_{\alpha,\alpha'}-1)\left|\bar{\alpha} \right>_{l,l+1}\left<\bar{\alpha'} \right| e^{-i\varphi_{l,\alpha,\alpha'}}+\text{H.c.} \}.\notag
\end{eqnarray}
 Thus we can adjust $p_{l,\alpha}, t_{l,\alpha,\alpha'}$, $\varphi_{l,\alpha,\alpha'}$, and $\omega_{l,\alpha,\alpha'}$ to implement different forms of spin-orbit coupling. The on-site potential and the hopping patterns of the Hamiltonian before and after the unitary transformation are shown in  Figs. \ref{figure2}(b) and \ref{figure2}(c), respectively.

According to Eq. (\ref{eq17}), we choose $p_{l,\uparrow}=h_z$,  $p_{l,\downarrow}=-h_z$,  $t_{l,\uparrow,\uparrow}=-t_{l,\downarrow,\downarrow}=t_z$,   $t_{l,\uparrow,\downarrow}=t_{l,\downarrow,\uparrow}=\Delta_0$,  $\varphi_{l,\uparrow,\uparrow}=\varphi_{l,\downarrow,\downarrow}=0$,
$\varphi_{l,\uparrow,\downarrow}=-\pi/2+\varphi$, and $\varphi_{l,\downarrow,\uparrow}=-\pi/2-\varphi$, and the Hamiltonian becomes the Hamiltonian in Eq. (\ref{eq4}) that we want to simulate. In this case, the four drive frequencies added by an external magnetic flux are
\begin{eqnarray}
J_l(t)&=&4t_z\cos(\omega_{l,\uparrow,\uparrow}t) -4t_z\cos(\omega_{l,\downarrow,\downarrow}t)\notag\\
&&+4\Delta_0\cos(\omega_{l,\uparrow,\downarrow}t-\frac{\pi}{2}+\varphi)\notag\\
&&+4\Delta_0\cos(\omega_{l,\downarrow,\uparrow}t-\frac{\pi}{2}-\varphi),
\label{eq2}
\end{eqnarray}
where
\begin{eqnarray}
\omega_{l,\uparrow,\uparrow}&=&E_{l,\uparrow}-E_{l+1,\uparrow},\notag\\
\omega_{l,\uparrow,\downarrow}&=&E_{l,\uparrow}-E_{l+1,\downarrow}-2h_z,\notag\\
\omega_{l,\downarrow,\uparrow}&=&E_{l,\downarrow}-E_{l+1,\uparrow}+2h_z,\notag\\
\omega_{l,\downarrow,\downarrow}&=&E_{l,\downarrow}-E_{l+1,\downarrow},
\label{eq3}
\end{eqnarray}
and $4t_z$ and $4\Delta_0$,  $\varphi$, and $2h_z$ are the amplitudes, phase, and detuning, respectively. This time-dependent coupling strength $J_l(t)$ can be realized by adding external magnetic fluxes with DC and AC components threading the SQUIDs \cite{Gu,Lei-induc3-prl}. The hopping strengths and hopping phases both can be controlled by the amplitudes and phases of the AC flux. We set $h_z/t_0=0.3$ and then the smallest frequency distance between each pair is nearly 20 times the effective hopping strengths $t_0$ and $\Delta_0$, so these four drive frequencies can achieve the corresponding four hoppings, as shown in Fig. \ref{figure2}(b).

It should be pointed out that the proposed superconducting circuit is the same as that in Ref. \cite{Gu}, but the external magnetic field threading through the SQUIDs is different. We can change the phase, drive detuning, and amplitude of the magnetic field in each SQUID to achieve the lattice on-site potential, and synthetic spin-orbit coupling of different phases.

\begin{figure}[tbp]		
\centering
\includegraphics[width=\columnwidth]{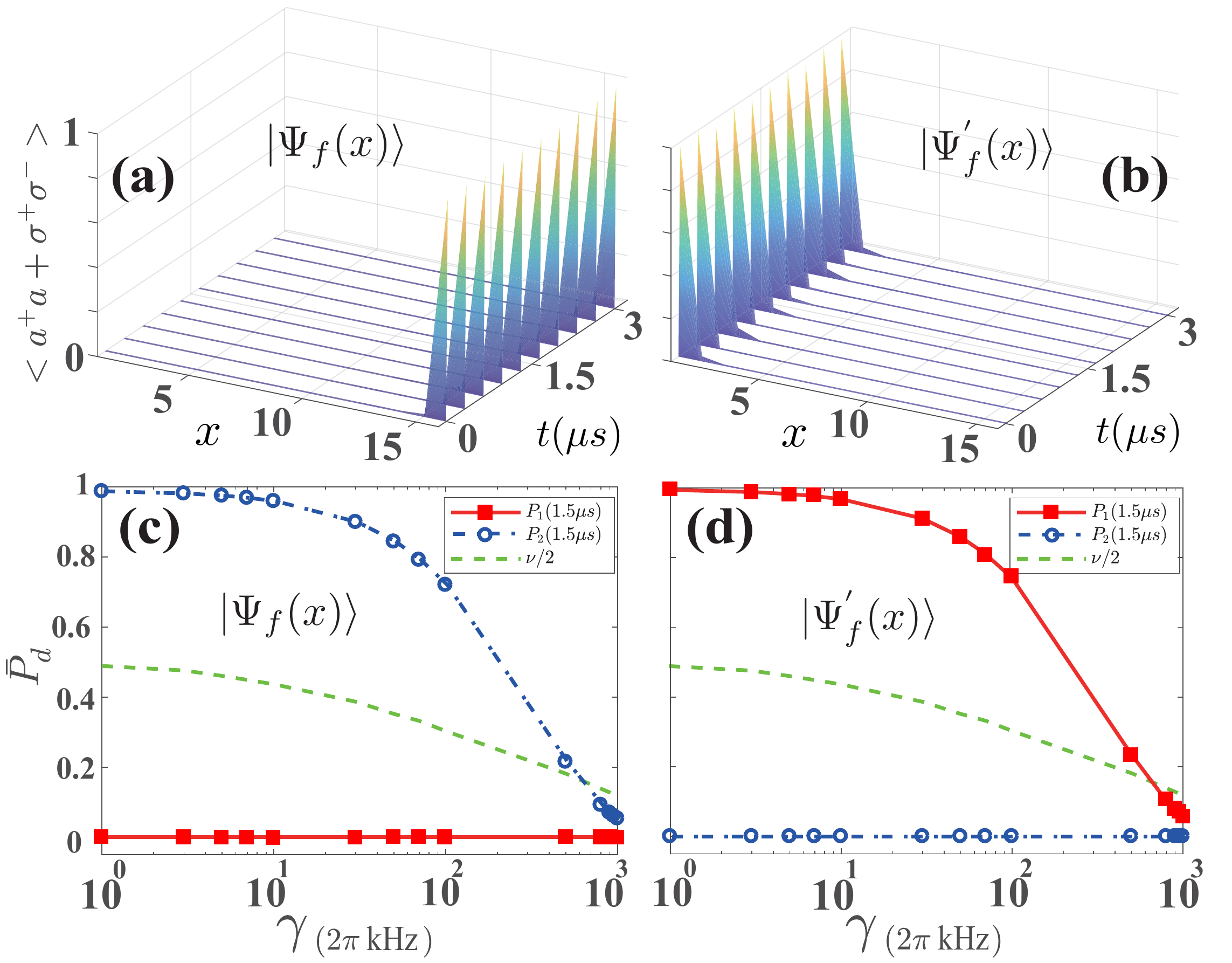}
\caption{Dynamical detection of polaritonic topological edge states. The time evolution of the polaritonic density distribution $\langle \sigma^+\sigma^- + \hat{a}^\dagger \hat{a}\rangle$ is shown when the JC lattice is in (a) $|\Psi_{f}(x)\rangle$, which is obtained by the $\hat{O}_1$ and $\hat{O}_2$ quantum operations, and (b) $|\Psi^{'}_{f}(x)\rangle$, obtained by the $\hat{O}_2$ and $\hat{O}_1$ quantum operations. Also shown are the  edge-site populations $P_1(t)$ and $P_2(t)$ at 1.5 $\mu$s and the oscillation center $\nu/2$ of edge states (c) $|\Psi_{f}(x)\rangle$ and (d) $|\Psi^{'}_{f}(x)\rangle$  for different decay rates $\gamma$.}
\label{figure3}
\end{figure}

\section{Detection of topological properties}

According to Eqs.~(\ref{eq7}) and (\ref{eq8}), or as shown in Figs. \ref{figure1}(c) and \ref{figure1}(d), the polariton in the left or right edge state is maximally distributed in the leftmost and rightmost JC lattice sites. Their internal spins are in the superposition states  $\left( |\uparrow\rangle_l+i|\downarrow\rangle_l \right)/\sqrt{2}$ and $\left(|\uparrow\rangle_l-i|\downarrow\rangle_l \right)/\sqrt{2}$, respectively.
In our demonstration of the non-Abelian statistics, the two final states $|\Psi_{f}(x)\rangle$ and $|\Psi^{'}_{f}(x)\rangle$ correspond to the two edge states, which will mostly be localized in their corresponding edge sites for a long time. Therefore, by detecting the population of the edge sites, we can successfully verify the final states.

Figure \ref{figure3}(a) shows the result of detecting the state $|\Psi_{f}(x)\rangle$ by applying first  $\hat{O}_1$ and then $\hat{O}_2$ quantum operations and Fig. \ref{figure3}(b) detection of  the state $|\Psi^{'}_{f}(x)\rangle$ obtained by $\hat{O}_2$ and $\hat{O}_1$. The initial states of the two detections are taken as
\[
\begin{aligned}
|\Psi_f(t=0)\rangle&=|0\text{g}\rangle_1\cdots|0\text{g}\rangle_{N-1} \left(|\uparrow\rangle_N-i|\downarrow\rangle_N \right)/\sqrt{2},\\
|\Psi^{'}_f(t=0)\rangle&=\left(|\uparrow\rangle_1+i|\downarrow\rangle_1 \right)|0\text{g}\rangle_2\cdots|0\text{g}\rangle_N/\sqrt{2}.\\
\end{aligned}
\]
It can be seen that after an evolution of 3 $\mu$s, because of topological protection, the final density distribution of the polaritons in the JC model lattice is still mostly distributed at the corresponding ends. Therefore, the two quantum states $|\Psi_{f}(x)\rangle$ and $|\Psi^{'}_{f}(x)\rangle$ are experimentally distinguishable.

The polaritonic topological winding number can be related to the time-averaged dynamical chiral center associated with the single-polariton dynamics \cite{Gu,Mei}, i.e.,
\begin{equation}
	\nu ={\lim_{T\rightarrow \infty }}\frac{2}{T}\int_{0}^{T}dt\, \langle \psi_{\text{c}}(t)| \hat{P}_{\text{d}} |\psi_{\text{c}}(t)\rangle,
    \label{eq10}
\end{equation}
where $T$ is the evolution time and $\hat{P}_{\text{d}}=\sum_{l=1}^{N}l \bm{\sigma_y}^l$,   $|\psi_{\text{c}}(t)\rangle =\exp(-iHt)|\psi_{\text{c}}(0)\rangle$ is the time evolution of the initial single-polariton state $|\psi_{\text{c}}(0)\rangle=|0\text{g}\rangle_1\cdots |\uparrow\rangle_{\lceil N/2 \rceil} \cdots|0\text{g}\rangle_N$, where one of the middle JC lattice site has been put one polariton in, with its spin prepared in the state $|$$\uparrow\rangle$.
In Fig. \ref{figure3}(c) and \ref{figure3}(d), we plot the edge-site population
\begin{equation}
\begin{aligned}
&P_1(t)=\text{Tr}\left[\rho(t)\left(a_1^\dag a_1+\sigma_1^+\sigma_1^-\right)\right],\\
&P_2(t)=\text{Tr}\left[\rho(t)\left(a_N^\dag a_N+\sigma_N^+\sigma_N^-\right)\right],
\end{aligned}
\end{equation}
after 1.5 $\mu$s and the oscillation center $\nu/2$ of the state $|\Psi_{f}(x)\rangle$ and $|\Psi^{'}_{f}(x)\rangle$ for different decay rates. It shows that the edge state population and the chiral center smoothly decrease when the decay rate increases. It can be seen that as the decay rate $\gamma$ continues to increase, it will run inside the system,   the edge state will disappear due to noise, and the detection fails.

Finally, the influence of the system noise on the photon number and the decoherence of the qubit is evaluated by numerically integrating the  Lindblad master equation, which can be written as
\begin{equation}
    \dot {\rho }=-{i}[H_{\text{JC}},\rho ]+\sum_{l=1}^{N}\sum _{i=1}^{3}\gamma\left(\Gamma_{l,i}\,\rho \Gamma_{l,i}^{\dagger }-{\frac {1}{2}}\left\{\Gamma_{l,i}^{\dagger }\Gamma_{l,i},\rho \right\}\right),
    \label{eq11}
\end{equation}
where $\rho$ is the density operator of the whole system, $\gamma$ is the decay rate or noise strength (which are set to be the same here), and $\Gamma_{l,1}=a_{l}$,  $\Gamma_{l,2}=\sigma^-_{l}$,  and $\Gamma_{l,3}=\sigma^z_{l}$ are the photon-loss, transmon-loss,  and transmon-dephasing operators in the $l$th lattice, respectively. The typical decay rate is $\gamma=2\pi\times5$ kHz; at this decay rate,
the  detection of the edge state $|\Psi_{f}(x)\rangle$ results in $P_1(\tau) = 0$ and $P_2(\tau)= 0.974$ when $\tau=1.5$ $\mu$s, which correspond to a chiral center $\nu/2\simeq 0.451$. For the edge state $|\Psi^{'}_{f}(x)\rangle$, we have $P_1(\tau) = 0.971$ and $P_2(\tau)= 0$, which correspond to a chiral center $\nu/2\simeq 0.453$. Because of topological protection, the system is less affected by decoherence effect and these data are sufficient to distinguish edge states $|\Psi_{f}(x)\rangle$ and $|\Psi^{'}_{f}(x)\rangle$.

\section{Conclusion}

We have proposed to establish a 1D chain with superconducting circuits and show that the non-Abelian statistics can be demonstrated experimentally.The advantages of a superconducting circuit system make our scenario more feasible and stable, which will facilitate research to achieve a quantum computer. In addition, we also discussed the effect of decoherence on the edge state of the system and the results prove that our protocol will stay reliable under decoherence, which is very important for realizing quantum computation in experiments.

\acknowledgements
This work was supported by the Key-Area Research and Development Program of GuangDong Province (Grant No. ~2018B030326001), the National Natural Science Foundation of China (Grants No. ~11704180, No. ~11874156, and No. ~11904111), the National Key R\&D Program of China (Grant No. ~2016YFA0301803), and the project funded by China Postdoctoral Science Foundation (Grant No. ~2019M652684).

\end{document}